\documentclass{article}

\PassOptionsToPackage{numbers,compress,sort}{natbib}
\usepackage[final]{nips_2017}

\usepackage[utf8]{inputenc} 
\usepackage[T1]{fontenc}    
\usepackage{hyperref}       
\usepackage{url}            
\usepackage{booktabs}       
\usepackage{amsfonts}       
\usepackage{nicefrac}       
\usepackage{microtype}      
\usepackage{xcolor}
\usepackage{amsmath}
\usepackage{subcaption}
\usepackage[misc]{ifsym}
\usepackage[noabbrev,capitalize]{cleveref}
\usepackage{graphicx}

\title{Measuring Territorial Control in Civil Wars\\Using Hidden Markov
  Models:\\A Data Informatics-Based Approach}

\author{
  Therese Anders\textsuperscript{(\Letter)}\quad Hong Xu\quad Cheng Cheng\quad T. K. Satish Kumar\\
  University of Southern California,  Los Angeles, CA 90089, USA \\
  \texttt{\{tanders, hongx, chen260\}@usc.edu}, \texttt{tkskwork@gmail.com}\\
}

\begin{document}

\maketitle

\begin{abstract}
  Territorial control is a key aspect shaping the dynamics of civil war.
  Despite its importance, we lack data on territorial control that are
  fine-grained enough to account for subnational spatio-temporal
  variation and that cover a large set of conflicts. To resolve this
  issue, we propose a theoretical model of the relationship between
  territorial control and tactical choice in civil war and outline how
  \textit{Hidden Markov Models (HMMs)} are suitable to capture
  theoretical intuitions and estimate levels of territorial control. We
  discuss challenges of using HMMs in this application and mitigation
  strategies for future work.
\end{abstract}

\section{Introduction}

Territorial control is a key aspect shaping the dynamics of civil war. Information about who commands what degree of control in which area is crucial for military and civilian actors on the ground, as well as analysts of the conflict. Territorial control allows armed actors to extract resources, increase their mobilization base, and pursue civilian collaboration. For development practitioners and scholars of civil war, territorial control is a central aspect in analyzing patterns of civilian victimization and the provision of public goods. Existing research suggests that when actors do not command high levels of control, atrocities against noncombatants become more likely~\cite{aronsonetal2017,schutte2017,stewartliou2017,kalyvas2006} and the civilian population is at a higher risk to suffer from an under-provision of public goods~\cite{stewartforthcoming}.

Despite its centrality for both practitioners and scholars working
within the realm of the security-development nexus, we lack data on
territorial control that are fine-grained enough to account for
subnational spatio-temporal variation and cover a large set of
conflicts
. Existing efforts of
measuring territorial control in civil wars are either too crude to
account for subnational spatio-temporal
variation~\cite{stewartliou2017,pologleditsch2016,delacallesanchezcuenca2015,cunninghametal2013,delacallesanchezcuenca2012},
reduce to simple center versus periphery dynamics~\cite{schutte2017},
are limited to a single conflict
setting~\cite{kalyvaskocher2009,kalyvas2006}, or are resource
intensive to create~\cite{aronsonetal2017,taoetal2016}. We fill this
void by proposing a novel measurement strategy that utilizes publicly
available event data and reasons with it in a machine learning framework
to estimate territorial control in asymmetric civil wars.

\begin{figure}[t]
  \centering
  \includegraphics[width=0.9\textwidth]{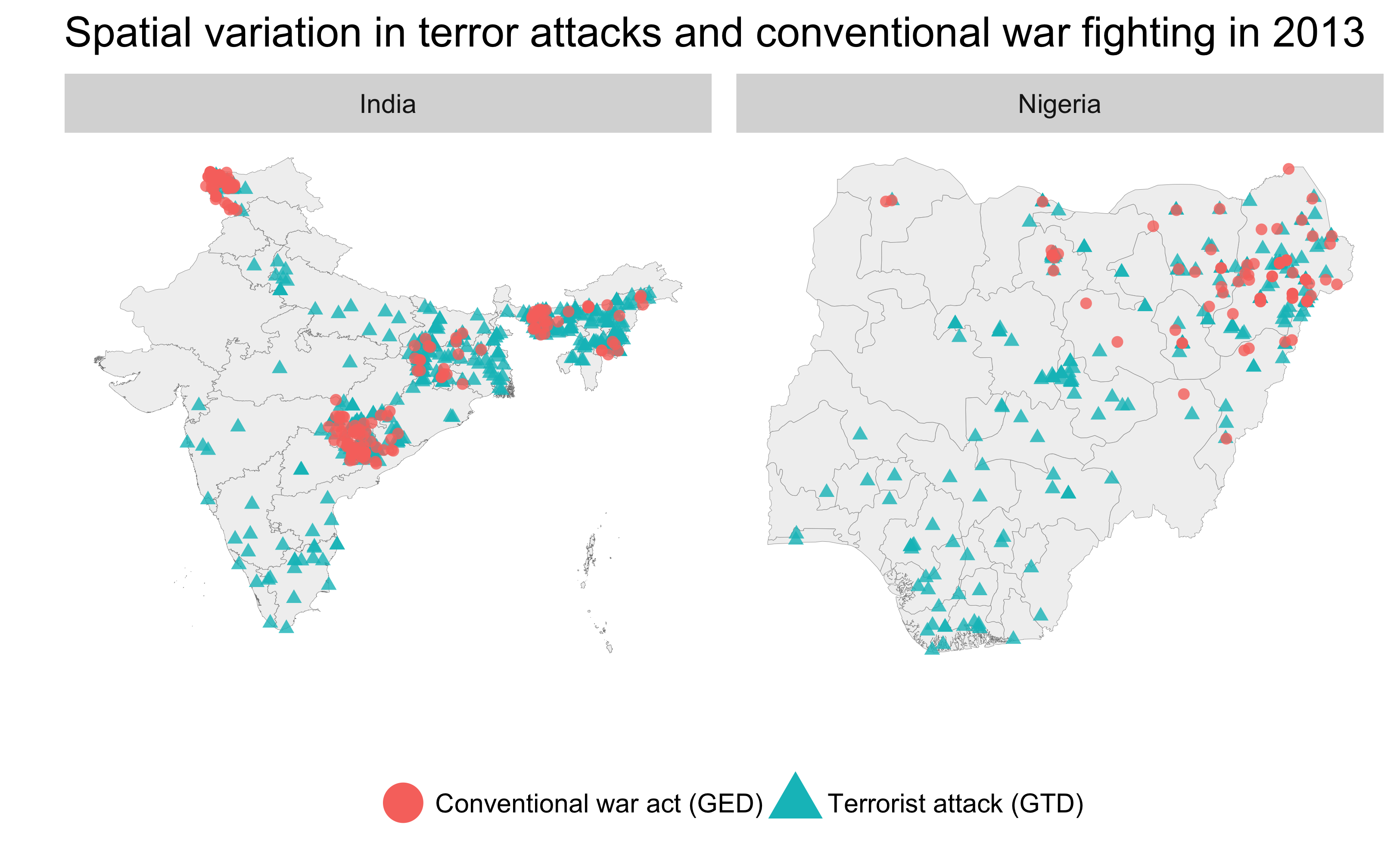}
  \caption{The plots illustrate the spatial variation in the
    co-occurrence of terrorist events and conventional war acts in India
    and Nigeria in 2013. Plotted in red dots are event locations from
    the UCDP Georeferenced Event Dataset (GED)~\cite{croicusundberg2017}
    and in blue triangles events from the Global Terrorism Database
    (GTD)~\cite{start2016}.\textsuperscript{*}}\label{fig:nigeria}

  \begin{flushleft}
    \small \textsuperscript{*} This figure consists of data from GED and
    GTD\@. Data from GED and GTD are owned by their respective owners
    and are distributed under other licenses. Shape files describing
    country borders come from the Global Administrative Areas version
    2.8 (\url{http://www.gadm.org}).
  \end{flushleft}
\end{figure}

Our approach leverages the insight that the level of rebel territorial
control influences their tactical choice. Broadly speaking, rebels
choose between fighting the government conventionally via direct
military confrontation, or coercively via terrorism against non-military
state forces and the civilian population. Indeed, as \cref{fig:nigeria} shows for India and Nigeria in 2013, we observe surprisingly little overlap between conventional war fighting and terrorist attacks.

Drawing from insights of the existing literature, we expect that the lower the level of
territorial control by rebels, the higher their propensity to use
terrorism will be, and vice versa~\cite{delacallesanchezcuenca2015,carter2015}.\footnote{A
  strong power asymmetry between the rebels and the government is a
  scope condition for the applicability of our model because only
  relatively weak rebels are likely to resort to terrorism. Based on existing
  estimates of rebel strength~\cite{cunninghametal2013}, and considering
  only cases in which the rebels were weaker or much weaker than the
  government, the method could be applied to approximately 50 cases
  between 1997 and 2014.} We
overlay our area of observation with a grid structure and conceptualize the number of terrorism and conventional war acts
within each grid cell as an observable indicator of an underlying unobserved variable territorial control.

Based on this theoretical model of the
relationship between territorial control and tactical choice in civil
war, in this extended abstract, we outline how a \textit{Hidden Markov
Model (HMM)} is suitable to capture theoretical intuitions
and estimate levels of territorial control within a machine learning
framework. We also discuss some challenges and mitigation strategies for
future work.

\section{Model and data}

We divide the region of interest into small cells and employ an
HMM~\cite{r89} for each of these cells. In an HMM, the \textit{true
  state} (also referred to as the \textit{hidden state}) of the system---here, the level of territorial control---is
hidden from observation and its evolution follows a Markov process. This
explicitly models the temporal dependencies of territorial control. Each
hidden state induces a probability distribution over possible
observable outputs.
In the Markov process, the probability of transitioning from one hidden
state to another is called a
\textit{transition probability}. The probability that a hidden state
produces a specific observable output is called an \textit{emission
  probability}.

In our HMM, each hidden state represents a level of government/rebel
territorial control of a specific geographical region.
Each hidden state at year \(t\) produces an observable output \(z_t\). Here, we use the
numbers of terrorist attacks (\(T\)) and conventional war
acts (\(C\)) as observable outputs. We aggregate the annual number of
terrorist attacks from the GTD~\cite{start2016}
and conventional war acts from the GED databases~\cite{croicusundberg2017} to 0.5-by-0.5 decimal degree cells
using the PRIO-GRID~\cite{tollefsenetal2015}.\footnote{We consider only
events that can at minimum be linked to a second order administrative region. For the GED, we consider only state-based events---excluding events that are coded as either violence against civilians or non-state conflicts. For the GTD, we exclude terrorist attacks perpetrated against military targets.}

To take the spatial correlation between cells into account, we employ a
\textit{Hidden Markov Random Field (HMRF)}. We consider the class of HMRFs that consist of
a collection of HMMs and assume correlation between the hidden states
of these HMMs. In the context of territorial control in civil wars, we
impose a correlation, such as the Potts model, only between HMMs of neighboring cells to model
the spatial correlation thereof.

\section{Discussion of challenges and mitigation strategies}

HMMs are well suited to model territorial control for two reasons.
First, we know that certain phenomena, such as tactical choice by
rebels, are likely to be highly correlated with, and their patterns
therefore informative about, levels of territorial control on the
ground~\cite{delacallesanchezcuenca2015,carter2015,delacallesanchezcuenca2012}.
HMMs allow us to model this intuition. If territorial control influences
tactical choice, we can perceive observed patterns of war fighting as an
emission of the underlying (unobserved) level of territorial control.
Second, since HMMs explicitly model the sequencing of hidden states,
they are a suitable tool to model the evolution of territorial control
over time.

However, the measurement of territorial control via HMMs also entails a
number of challenges. First, HMMs are typically used to model a single
sequence of states. However, in our application domain, the evolutions
of territorial control between neighboring cells are unlikely to be
independent. Diffusion processes for both the location and escalation of
violence are well documented~\cite{schutteweidmann2011}. Although
conceptually it is easy to incorporate spatial dependency into our
model, it introduces disproportionate computational difficulties. An HMM
can be solved by using fast and simple algorithms (such as the
Baum-Welch algorithm~\cite{r89}), but an HMRF usually requires much more
complicated approaches. For example, a method utilizing \textit{Markov
  Chain Monte Carlo} is known to solve an HMRF that may suit the
characteristics in our application~\cite{gr02}.

Second, even in the case of very active insurgencies, terrorist
attacks or conventional war as captured by widely used event datasets are
relatively rare events and therefore not normally distributed. This forces us to assume the number of events to be drawn from a Poisson distribution, rather than a Gaussian distribution, which is usually more difficult to analyze.
Making
things worse, there is a trade-off between the number of events per unit
of observation and the cell size of the imposed grid structure. The more
fine-grained the grid structure (and therefore the higher the resolution of
the resulting estimates of territorial control), the lower the number of
events that are observed per grid cell will be. Therefore, we need to estimate the optimal size of grid cells. We implement this by varying the cell size and shape (squares versus hexagons) using simulated data.

Third, standard HMMs do not allow for the inclusion of covariates. However,
a number of available data are informative about an
area's underlying propensity to change between different states of territorial control, such as land cover or
terrain. One avenue to leverage covariates might be to model the transition and emission probabilities to be perturbed by functions of these factors. For example, dense forest cover might make it harder for the government to re-capture rebel strongholds---thus the probability of transiting between different states could be subtracted by a monotonically increasing function of forestation.

Furthermore, standard HMMs assume the hidden states to be discrete.
However, while we can conceptualize territorial control to fall into
discrete categories, in reality it is likely to be a continuous measure
on a spectrum from full rebel to full government control. It is well
known that modeling HMMs with continuous hidden states is in general
much harder than that with discrete hidden states. One direction to
resolve this issue may be to apply a Kalman or particle filter.

Finally, the motivation of this paper lies in the scarcity of existing
data on territorial control that is fine-grained enough to account for
temporal and spatial variation on the subnational level. While this
increases the marginal value of new estimates, it also means that
appropriate training and testing data for our algorithm is hard to
obtain. To resolve this issue, we consider simulated data leveraging
insights from well-established theories of territorial
control~\cite{delacallesanchezcuenca2015,kalyvas2006}.

\section{Conclusion}

In this extended abstract, we proposed that the HMM and its extensions
are suitable for estimating general patterns of territorial control in
asymmetric civil wars. Future work tackles the challenge of spatial
correlation between HMMs of neighboring cells by using the HMRF and
developing better algorithms to account for continuous hidden states and
any other spatial dependencies on domain-specific factors, such as land
cover and terrain.

\bibliographystyle{abbrvnat}
\bibliography{refs}

\end{document}